\documentclass[conference,10pt]{IEEEtran}
\usepackage{epsfig,epsf,rotating,setspace,latexsym,amsmath,amssymb,amsfonts,bm,theorem,subfigure,epstopdf}
\usepackage{cite,authblk}
\usepackage{bbm}
\usepackage{algorithm}
\usepackage[noend]{algpseudocode}
\usepackage{color}
\usepackage{mathtools}
\usepackage{soul}

\algrenewcommand\algorithmicforall{\textbf{foreach}}
\algrenewcommand\algorithmicindent{.8em}

\newtheorem{theorem}{Theorem}
\newtheorem{lemma}{Lemma}

\newenvironment{Proof}[1]{\medskip\par\noindent{\bf Proof:\,}\,#1}{{\mbox{\,$\blacksquare$}\par}}

\IEEEoverridecommandlockouts
\allowdisplaybreaks

\begin{document}
 
\title{Timely Asynchronous Hierarchical Federated Learning: Age of Convergence}
 
\author{Purbesh Mitra \qquad Sennur Ulukus\\
        \normalsize Department of Electrical and Computer Engineering\\
        \normalsize University of Maryland, College Park, MD 20742\\
        \normalsize  \emph{pmitra@umd.edu} \qquad \emph{ulukus@umd.edu}}
\maketitle

\begin{abstract}
We consider an asynchronous hierarchical federated learning (AHFL) setting with a client-edge-cloud framework. The clients exchange the trained parameters with their corresponding edge servers, which update the locally aggregated model. This model is then transmitted to all the clients in the local cluster. The edge servers communicate to the central cloud server for global model aggregation. The goal of each client is to converge to the global model, while maintaining timeliness of the clients, i.e., having optimum training iteration time. We investigate the convergence criteria for such a system with dense clusters. Our analysis shows that for a system of $n$ clients with fixed average timeliness, the convergence in finite time is probabilistically guaranteed, if the nodes are divided into $O(1)$ number of clusters, that is, if the system is built as a sparse set of edge servers with dense client bases each.
\end{abstract}

\section{Introduction}\label{section: introduction}

Federated learning (FL), since its introduction in \cite{McMahan2016CommunicationEfficientLO}, has been studied extensively in distributed learning frameworks. Distributed learning (DL), as opposed to centralized learning, is often a useful tool in various applications where factors such as distributed data storage and data privacy are of concern, see e.g., \cite{McMahan2016CommunicationEfficientLO, Feng_mobility, Damaskinos_online_learning, Li_healthcare}. The FL model used in these applications consists of a central parameter server (PS) and multiple clients which have the distributed data. The PS has a global model which has to be trained on all the client data. In the FL settings, the PS sends copies of the global model to individual clients, which are then trained locally. After training, the clients send back their individual local models or their gradients, which are securely aggregated at the PS by weighted averaging. In \cite{McMahan2016CommunicationEfficientLO}, the authors showed that this training method yields convergence to a local optimum.

Due to the distributed nature of FL, its convergence is affected by factors such as data heterogeneity and communication overhead \cite{kairouz2021FLreview, shome2022federated}. The convergence result in \cite{McMahan2016CommunicationEfficientLO} relies on idealistic assumptions, such as, that the data is distributed in i.i.d.~manner across all clients, and that the client devices participate in the federated training with regularity. In reality, the distributed data is often non-i.i.d.~and the client devices are often unavailable in training cycles due to inefficient communication channels, energy availability of wireless devices, and the presence of Byzantine stragglers. To circumvent these issues, several solutions have been proposed in the literature, such as, priority-based client selection \cite{perazzone2022communication}, clustering \cite{briggs2020federated, wang2022demystifying}, adaptive quantization \cite{mao2022communication}, multi-task learning \cite{li2020federated, karimireddy2020scaffold, mortaheb2022fedgradnorm, vahapoglu2022fedgradnorm, mortaheb2022personalized, fallah2020personalized, wang2022federated}, asynchronous FL \cite{xie2020asynchronous}, and hierarchical FL \cite{liu_HFL}. 

\begin{figure}[t]
\centerline{\includegraphics[scale=0.6]{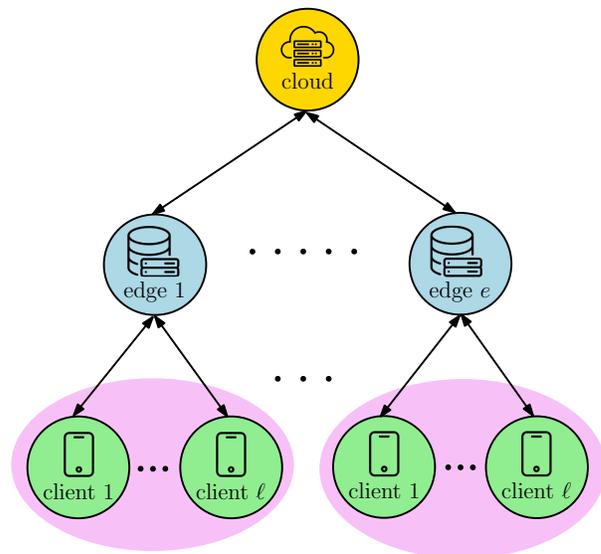}}
\caption{Hierarchical federated learning (HFL) with client-edge-cloud setting.}
\label{fig: HFL}
\vspace*{-0.4cm}
\end{figure}

In particular, the client-edge-cloud based hierarchical FL (HFL) setting, introduced in \cite{liu_HFL}, is used for circumventing the problem of heterogeneity with provable guarantees \cite{abad_HFL}. In this HFL setting, there are multiple clusters of clients which are connected to their corresponding edge servers, which themselves are connected to a central cloud server. The edge servers perform local model aggregation from their associated client devices, which leads to a convergence of the local models. For global convergence, the edge servers update the cloud server, typically less frequently than the clients, to get back a globally aggregated model, which is then passed on to the client devices. Several variants of this model, such as, asynchronous hierarchical FL (AHFL) \cite{wang2022asynchronous, Chen_shfl}, have been proposed in the literature which enable the training process to converge without any synchronicity requirement. 

\begin{figure*}[t]
\centerline{\includegraphics[scale=0.8]{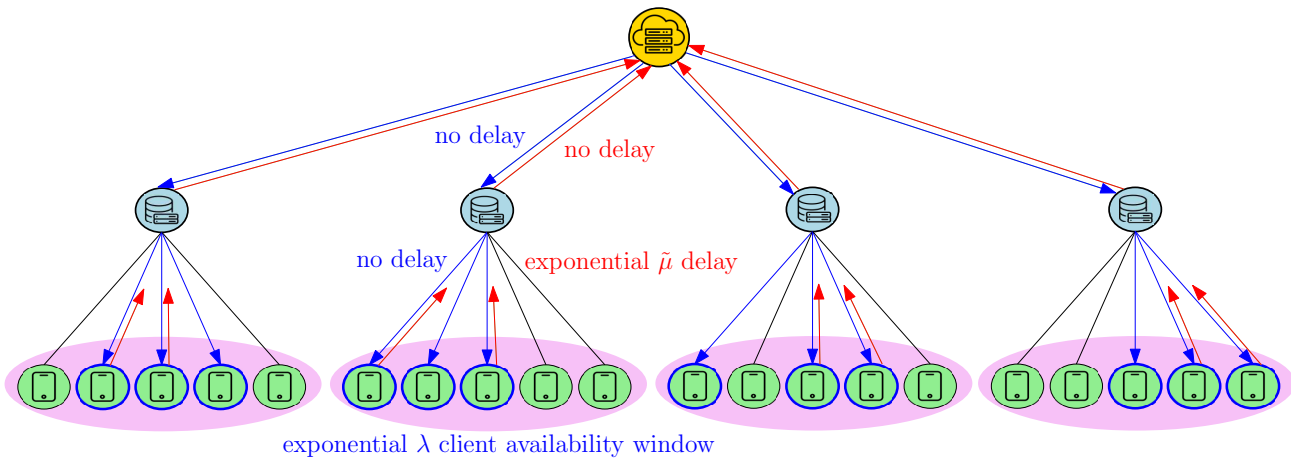}}
\caption{Timely asynchronous hierarchical federated learning with client-edge-cloud setting. The number of edge servers $e=4$. Each edge server has $\ell=5$ clients associated with it. At each training iteration, edge server waits for $m=3$ clients to be available, and only the first $k=2$ earliest finishers are aggregated at the edge. The local aggregation is then globally aggregated at the cloud asynchronously.}
\label{fig: TAHFL}
\vspace*{-0.3cm}
\end{figure*}

In all of these FL settings, better communication between clients and the PS is necessary. Works in \cite{perazzone2022communication, Zhao2018FederatedLW, Nishio2018ClientSF, Amiri_DSGD, Barnes2020rTopkAS,  Dhakal_codedFL, Amiri_FL_update, chang2020communication, hamer2020fedboost, hyeon_fedpara, qiao2021communication, vithana2022fsl, liu2022time, ali2023delay, wen2022federated, Bouacida_dropout, chen2022fedobd, yang2020age} have proposed various schemes, which yield faster and reliable communication. Another criteria for efficient communication is better timeliness of the system, which makes the overall training process faster. This aspect is investigated in \cite{buyukates_timely_FL}, where the authors have shown that selecting any random $k$ out of $n$ clients as in \cite{McMahan2016CommunicationEfficientLO} does not necessarily yield a fast convergence in FL. \cite{buyukates_timely_FL} uses the age of information metric \cite{kosta17AoIbook, Sun2019AgeOI, yatesJSACsurvey, melih_cache_TWT, kaswan2021ISIT}, a widely used metric in the communications and networking literature, to characterize the timeliness of the individual clients. \cite{buyukates_timely_FL} also proposes a timely FL scheme, where the PS waits for $m$ out of the $n$ clients to be available and then sends the updated model. Then, for model aggregation, only the first $k$ responses out of the $m$ clients are chosen. \cite{buyukates_timely_FL} shows that this scheme results in faster training cycles, and hence faster convergence. This gain in performance is due to the fact that, in the proposed scheme in \cite{buyukates_timely_FL}, the initial waiting period ensures that all the clients are available in the training as opposed to the random selection scheme in \cite{McMahan2016CommunicationEfficientLO}.

Motivated by this approach, in this work, we focus on the timeliness of a more realistic FL setting, namely AHFL. In particular, we consider the client-edge-cloud setting, where the local client-edge training happens in the same way as in \cite{buyukates_timely_FL} and the edge-cloud aggregation takes place asynchronously. The cloud model updates its version number by one whenever it performs aggregation. On the client's side, if a client manages to send its trained model to the edge, and therefore, to the cloud, it catches up with the global version of the cloud; else, its version does not change. A client with its model being successfully aggregated in a training cycle will have the most recent version number, while a client trained on an older version incurs some \textit{staleness} in the system, which is denoted as the difference of versions between the cloud and the client. The analysis in \cite{xie2020asynchronous} shows that the global convergence of an asynchronous FL system is dominated by this staleness factor, i.e., the smaller is the staleness, the faster the system will converge. Since we are not considering a traditional FL setting, rather an AHFL system, we have to incorporate both the timeliness and the bounded staleness in the design of the system for better performance.

With this backdrop, in this work, we investigate the convergence criteria of a timely AHFL system consisting of dense clients. We show that there can be a probabilistically guaranteed global convergence if the number of clusters of the hierarchical system is $O(1)$, while maintaining timeliness. That is, a good operating setting for AHFL is to have sparse edge devices with dense clients.

\section{System Model}\label{section: system model}

In a FL setting, the parameters $\boldsymbol{\theta}\in\mathbb{R}^d$ of a model are trained in $n$ different client devices with distributed training data. The goal of the system is to minimize a loss function,
\begin{align}
    \mathcal{L}(\boldsymbol{\theta};D)=\frac{1}{n}\sum_{i=1}^{n}\mathcal{L}(\boldsymbol{\theta};D_i),
\end{align}
where $D_i$ is the training data available to the $i$th client and $D=\cup_{i=1}^{n} D_i$ is the total data set. In the beginning of the $t$th cycle, the PS sends a global model $\boldsymbol{\theta}_t$ to the clients. At each cycle, the clients perform $\tilde{t}\in\mathbb{N}$ step stochastic gradient descent (SGD) on their local objective function as follows
\begin{align}
    \boldsymbol{\theta}^{i}_{j+1}=\boldsymbol{\theta}^{i}_{j}-\eta^{i}_{j}\nabla \mathcal{L}(\boldsymbol{\theta}^{i}_{j};D_i),\quad j\in \left[\tilde{t}\right],i\in [n], 
\end{align}
where $\eta^{i}_{\ell}$ is the learning rate corresponding to the $j$th local step of the $i$th client and $\boldsymbol{\theta}^{i}_{1}=\boldsymbol{\theta}_t$. After training completion, each client sends the PS its model, which are aggregated as 
\begin{align}
    \boldsymbol{\theta}_{t+1}=\sum_{i=1}^{n}\frac{|D_i|}{|D|}\boldsymbol{\theta}^{i}_{\tilde{t}+1}.
\end{align}
This updated model is then used for the next training cycle and the process continues until the learning model converges.

In a HFL, the most commonly used model is the client-edge-cloud model, introduced in \cite{liu_HFL}. In this model, shown in Fig.~\ref{fig: HFL}, there are total $n$ clients distributed under $e$ edge servers. For simplicity, we assume that the clusters are symmetric and there are $\ell$ clients corresponding to each edge server, i.e., $n=e\cdot \ell$. For this work, we consider that the edge-client network is performing local update in a timely manner as described in \cite{buyukates_timely_FL}. This model is shown in Fig.~\ref{fig: TAHFL}. The clients are assumed to be available at each training cycle after an exponentially distributed time window with parameter $\lambda$. The first $m$ available clients among $\ell$ clients receive the model from the edge servers without any downlink communication delay. For any client, the training takes $c$ units of time, which is a deterministic quantity. The uplink communication to the edge is modeled as an exponential random variable with parameter $\tilde{\mu}$. For aggregation, only the first $k$ responding clients are considered. Analysis of \cite{buyukates_timely_FL} shows that choosing the $(k,m)$ pair with a linear dependency on $n$, i.e., $k=\alpha m$ and $m=\beta\ell$ are sufficient for maintaining an average timeliness independent of $\ell$, where $\alpha,\beta\in(0,1)$ are constants, which can be chosen optimally. 

\begin{algorithm}[t]
  \caption{Timely AHFL algorithm}
  \label{algo: AHFL}
  \begin{algorithmic}[1]
    \State Initialize $\boldsymbol{\theta}_0$ and send to all the clients.
    \For{$t\in[T]$}
        \Procedure{CloudUpdate}{}
          \State Receive $\boldsymbol{\theta}'$ from the edge server $s$.
          \State Perform global aggregation as $$\boldsymbol{\theta}^{t}\leftarrow(1-\sigma(t-t'))\boldsymbol{\theta}^{t-1}+\sigma(t-t')\boldsymbol{\theta}'.$$
          \State Send $\boldsymbol{\theta}^{t}$ to the edge server $s$.
        \EndProcedure
        \For{$s\in[e]$}
            \Procedure{EdgeUpdate}{edge $s$}
                \State Receive $\boldsymbol{\theta}_t$ from the cloud server.
                \State Start the client training cycle.
                \State Wait till $m$ among the $\ell$ clients to become available.
                \State Send $\boldsymbol{\theta}_t$ to the clients.
                \State Wait until the first $k$ clients (set $\mathcal{K}$) respond.
                \State Aggregate as $\boldsymbol{\theta}'=\sum_{i\in\mathcal{K}}\frac{|D_i|}{|D_{\mathcal{K}}|}\boldsymbol{\theta}^{i}_{\tilde{t}+1}$.
                \State Send $\boldsymbol{\theta}'$ to the cloud server.
            \EndProcedure
            \For{$i\in[\ell]$}
                \Procedure{ClientTraining}{client $i$}
                    \State Become available after exponential ($\lambda$) time.
                    \State Receive $\boldsymbol{\theta}_t$ from the edge.
                    \State $\boldsymbol{\theta}^{i}_{1}\leftarrow\boldsymbol{\theta}_t$.
                    \For{$j\in\left[\tilde{t}\right]$}
                        \State $\boldsymbol{\theta}^{i}_{j+1}=\boldsymbol{\theta}^{i}_{j}-\eta^{i}_{j}\nabla \tilde{\mathcal{L}}(\boldsymbol{\theta}^{i}_{j};D_i)$.
                    \EndFor
                    \State Send $\boldsymbol{\theta}^{i}_{\tilde{t}+1}$ to the edge in exponential ($\tilde{\mu}$) time.
                \EndProcedure
            \EndFor
        \EndFor
    \EndFor
  \end{algorithmic}
\end{algorithm}

In the timely AHFL system, the edge-cloud network, on the other hand, is doing global updates asynchronously with negligible delay due to their high data rate. To penalize the staleness, instead of doing SGD on $\mathcal{L}(\boldsymbol{\theta};D_i)$, each client performs SGD on the modified loss function
\begin{align}\label{modified_loss_function}
    \tilde{\mathcal{L}}(\boldsymbol{\theta};D_i)=\mathcal{L}(\boldsymbol{\theta};D_i)+\frac{\rho}{2}||\theta-\theta_t||^2
\end{align}
and sends it to the edge. Each time an edge server generates a local update, it immediately sends it to the cloud, which updates the global model by taking a weighted sum of the arrived local model $\boldsymbol{\theta}'$ and the existing global model $\boldsymbol{\theta}^{t}$,
\begin{align}\label{cloud_update_equation}
    \boldsymbol{\theta}^{t+1}=(1-\sigma(t-t'))\boldsymbol{\theta}^{t}+\sigma(t-t')\boldsymbol{\theta}',
\end{align}
where $t'$ is the last time-stamp of the trained model from the particular edge and $\sigma(\cdot)\in(0,1)$ is a decreasing function of the staleness, chosen optimally. This new model is then sent back to the edge servers, which are then used for client training in the subsequent cycle. The AHFL procedure is shown in Algorithm~\ref{algo: AHFL}.

We denote the training time duration of the $t$th cycle as $Y_t$ and the time duration of the trained model at the $i$th node to be aggregated as $Y^i_j$, where $j\in\mathbb{N}$ denotes the index of successful participation of node $i$. From the analysis of \cite{buyukates_timely_FL}, we obtain the expression of an average training cycle as
\begin{align}\label{avg_training_cycle}
    \mathbb{E}[Y_t]=\mathbb{E}[Z_{m:\ell}]+c+\mathbb{E}[\tilde{X}_{k:m}],
\end{align}
where $Z_{m:\ell}$ denotes the time it takes for $m$ out of $\ell$ clients to be available. Using the order statistics formulation in \cite{david2003order}, we obtain 
\begin{align}
    \mathbb{E}[Z_{m:\ell}]=\frac{1}{\lambda}(H_{\ell}-H_{\ell-m}),
\end{align}
where $H_{\ell}=\sum_{j=1}^{\ell}\frac{1}{j}$. Further, $\tilde{X}_{k:m}$ in \eqref{avg_training_cycle} denotes the time for the first $k$ out of $m$ clients to send their model to the PS. Using similar formulation, we can write
\begin{align}
    \mathbb{E}[\tilde{X}_{k:m}]=\frac{1}{\tilde{\mu}}(H_{m}-H_{m-k}).
\end{align}
Due to the symmetry of the network, the probability of a particular client being available for training and successfully sending its trained model to the PS is $\frac{k}{\ell}$. Therefore, we obtain the following expectation
\begin{align}\label{avg_training_of_i}
    \mathbb{E}[Y^i_j]=\frac{\ell}{k}\mathbb{E}[Y_t].
\end{align}
Since the edge-cloud communication is asynchronous and instantaneous, the expectation in \eqref{avg_training_of_i} also implies the expected time of the $i$th client's version number to be updated.

\begin{figure*}[t]
\centerline{\includegraphics[width=0.7\textwidth]{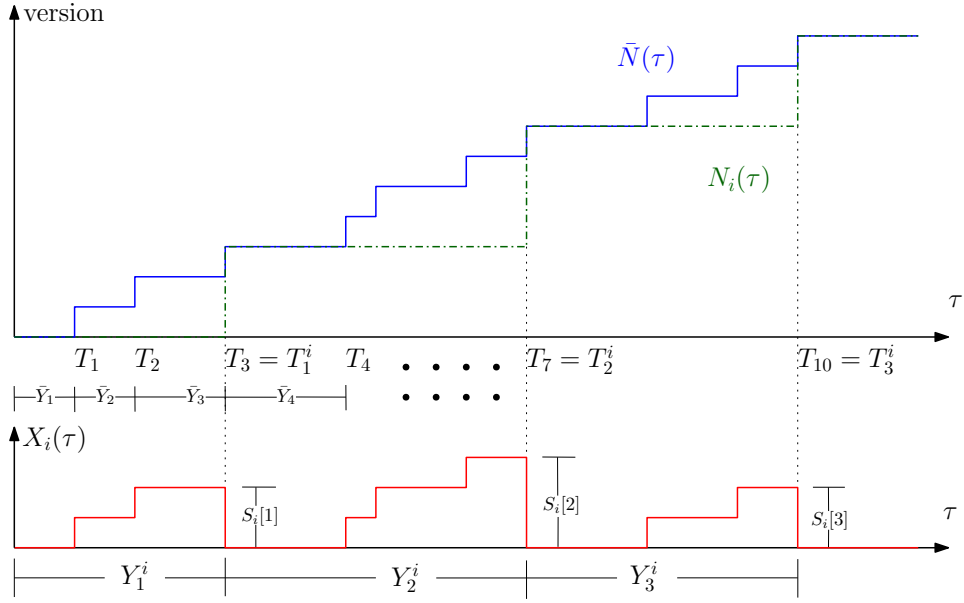}}
\caption{Sample paths of version evolution in the cloud server and the $i$th node and the corresponding staleness.}
\label{fig: version}
\end{figure*}

\section{Convergence Guarantee of AHFL}\label{section: convergence}

In this section, we formulate the expression for the version difference between the cloud and a client, and derive an upper bound. We denote $\Bar{N}(\tau)$ and $N_i(\tau)$ as the counting processes of the cloud version and the $i$th node version, respectively, as shown in Fig.~\ref{fig: version}. The process $\Bar{N}(\tau)$ is increasing at every $T_j,j\in\mathbb{N}$, with $\Bar{Y}_j$ being the inter-update times, whereas the process $N_i(\tau)$ gets updated at every $T^i_j,j\in\mathbb{N}$, with mean inter-arrival time $\mathbb{E}[Y^i_j]$. Using a similar formulation as \cite{melih2020infocom}, we define the version difference for the $i$th node as
\begin{align}\label{version difference}
    X_i(\tau)=\Bar{N}(\tau)-N_i(\tau).
\end{align}
Since $N(\tau)$ always leads ahead of $N_i(\tau)$, $X_i(\tau)$ is always non-negative. From \eqref{version difference}, we can write the staleness of the $j$th successful training of the $i$th node as
\begin{align}\label{staleness}
    S_i[j]=X_i(T^{i-}_j), \quad\forall j\in\mathbb{N}.
\end{align}

Now, in the convergence analysis of \cite{xie2020asynchronous}, the authors have shown that for asynchronous federated optimization with $T$ training cycles and the condition that $S_i[j]\leq M,\forall j\in[T]$,
\begin{align}
    \min_{t=0,\ldots,T-1}\mathbb{E}\left[||\nabla\mathcal{L}(\boldsymbol{\theta}_t;\mathcal{D})||^2\right]\leq O(M^2).
\end{align}
The same criterion also holds true for our AHFL system. Therefore, to have faster convergence, it is necessary to bound the staleness tightly. Since we have defined $\{S_i[j]\}_{j\in\mathbb{N}}$ as a random process in \eqref{staleness}, we use the Markov inequality to obtain the following bound
\begin{align}\label{bounded_prob}
    \mathbb{P}\big(S_i[j]\leq M\big)\geq 1-\frac{\mathbb{E}\left[S_i[j]\right]}{M}.
\end{align}
From \eqref{bounded_prob}, we conclude that minimizing $\mathbb{E}\left[S_i[j]\right]$, would result in tighter upper bound. In the following lemma, we derive the expression of the expectation of staleness.

\begin{lemma}\label{lemma: expected_staleness}
    The expected staleness of an AHFL system in steady state is given by
    \begin{align}
        \lim_{j\to\infty}\mathbb{E}\left[S_i[j]\right]=\frac{\mathbb{E}[Y^i_j]}{\mathbb{E}[\Bar{Y}]}-1.
    \end{align}
\end{lemma}

\begin{Proof}
We can write the staleness as difference of the counting processes as following
\begin{align}\label{diff_count}
    S_i[j]=\Bar{N}(T^{i-}_j)-N_i(T^{i-}_j)
\end{align}
Since the version of $i$ does not change for $Y^i_j$ duration, we have $\Bar{N}(T^{i-}_j)=\Bar{N}(T^{i}_{j-1}+Y^i_j)-1$ and $N_i(T^{i-}_j)=\Bar{N}(T^i_{j-1})$. Substituting these values in \eqref{diff_count}, we obtain
\begin{align}\label{diff_renewal}
    S_i[j]=\Bar{N}(T^i_{j-1}+Y^i_j)-\Bar{N}(T^i_{j-1})-1.
\end{align}
Asymptotically, for large training time, we have $T^i_{j-1}\to\infty$ as $j\to\infty$. Now, since $Y^i_j$ is a random variable, we use the result from \cite[Lem.~2]{kaswan2023timely} (see also \cite{kaswan2023age_non_poisson}) and apply the modified Blackwell renewal theorem to evaluate the expectation of $S_i[j]$ from \eqref{diff_renewal} as follows
\begin{align}
    \lim_{j\to\infty}S_i[j]&=\lim_{T^i_{j-1}\to\infty}\Bar{N}(T^i_{j-1}+Y^i_j)-\Bar{N}(T^i_{j-1})-1 \\
    &=\frac{\lim_{j\to\infty}\mathbb{E}[Y^i_j]}{\mathbb{E}[\Bar{Y}]}-1=\frac{\mathbb{E}[Y^i_j]}{\mathbb{E}[\Bar{Y}]}-1.
\end{align}
This concludes the proof of the lemma.
\end{Proof}

We have the expression of $\mathbb{E}[Y^i_j]$ from Section~\ref{section: system model}. We evaluate the expectation of $\mathbb{E}[\Bar{Y}]$ in the following lemma.

\begin{lemma}\label{lemma: inter-update time}
    For a symmetric AHFL system with $e$ edge servers, the mean inter-update time of the cloud server is given by
    \begin{align}
        \frac{1}{\mathbb{E}[\Bar{Y}]}=\frac{e}{\mathbb{E}[Y_t]}.
    \end{align}
\end{lemma}

\begin{Proof}
We can write the renewal process $\Bar{N}(\tau)$ as a superposition of multiple renewal processes as follows
\begin{align}
    \Bar{N}(\tau)=\sum_{j=1}^{e}\Bar{N}_{j}(\tau),
\end{align}
where $\Bar{N}_j(\tau)$ is a renewal process which indicates the version number of the $j$th edge server $s_j$, which can be expressed as
\begin{align}
    \Bar{N}_j(\tau)=\max\{N_i(\tau):i\in s_j\}.
\end{align}
The cluster version updates after completion of a training cycle, i.e., with  inter-update time $Y_t$. For the steady state, i.e., for $\tau\to\infty$, we can write the following relation
\begin{align}\label{ss_avg}
    \lim_{\tau\to\infty}\frac{\Bar{N}(\tau)}{\tau}=\sum_{j=1}^{e}\lim_{\tau\to\infty}\frac{\Bar{N}_{j}(\tau)}{\tau}.
\end{align}
Using the renewal theorem, we obtain $\lim_{\tau\to\infty}\frac{\Bar{N}(\tau)}{\tau}=\frac{1}{\mathbb{E}[\Bar{Y}]}$ and $\lim_{\tau\to\infty}\frac{\Bar{N}_{j}(\tau)}{\tau}=\frac{1}{\mathbb{E}[Y_t]}$. Substituting these in \eqref{ss_avg} yields
\begin{align}
    \frac{1}{\mathbb{E}[\Bar{Y}]}=\sum_{j=1}^{e}\frac{1}{\mathbb{E}[Y_t]}=\frac{e}{\mathbb{E}[Y_t]}.
\end{align}
This concludes the proof of the lemma.
\end{Proof}

Now, using Lemmas~\ref{lemma: expected_staleness} and \ref{lemma: inter-update time}, we find the criteria for a probabilistic upper bound of the staleness of a node being finite in the following theorem. 

\begin{theorem}\label{thm: upper bound}
In a timely AHFL system with dense clients setting, if the number of clusters $e=O(1)$, then for any $0<\varepsilon<1$, there exists a constant $M$, such that the staleness of a node is bounded by $M$ with probability $\mathbb{P}\big(S^i_j\leq M\big)\geq 1-\varepsilon$.
\end{theorem}

\begin{Proof}
Using \eqref{avg_training_of_i} and substituting the result of Lemma~\ref{lemma: expected_staleness} into Lemma~\ref{lemma: inter-update time}, we obtain the expression for the expected staleness in steady state as
\begin{align}\label{expected staleness}
    \lim_{j\to\infty}\mathbb{E}\left[S_i[j]\right]&=\frac{\ell}{k}\mathbb{E}[Y_t]\times\frac{e}{\mathbb{E}[Y_t]}-1\\
    &=\frac{e\cdot\ell}{k}-1=\frac{n}{k}-1.
\end{align}
From the timeliness condition of \cite{buyukates_timely_FL}, $k=\alpha\beta\ell$. Substituting this in \eqref{expected staleness}, we obtain
\begin{align}\label{final staleness}
    \lim_{j\to\infty}\mathbb{E}\left[S_i[j]\right]=\frac{n}{\alpha\beta\ell}-1=\frac{e}{\alpha\beta}-1.
\end{align}
Now, using \eqref{final staleness} in \eqref{bounded_prob}, we obtain the following probabilistic bound
\begin{align}\label{bound on staleness}
    \mathbb{P}\big(S_i[j]\leq M\big)\geq 1-\frac{1}{M}\left(\frac{e}{\alpha\beta}-1\right).
\end{align}
Now, for a dense client setting, i.e., $n\to\infty$ and $e=O(1)$, the probabilistic bound in \eqref{bound on staleness} can be written as
\begin{align}\label{final bound}
    \lim_{n\to\infty}\mathbb{P}(S_i[j]\leq M)\geq 1-\lim_{n\to\infty}\frac{O(1)}{M}.
\end{align}
Clearly, there exists a finite $M$ for which the bound in \eqref{final bound} is satisfied asymptotically. Choosing any $M\geq\frac{1}{\varepsilon}\left(\frac{e}{\alpha\beta}-1\right)$ yields the statement of the theorem.
\end{Proof}

\begin{figure*}[t]
\subfigure[Loss vs. epochs for $n=100$.]{
\centering
\includegraphics[scale=0.55]{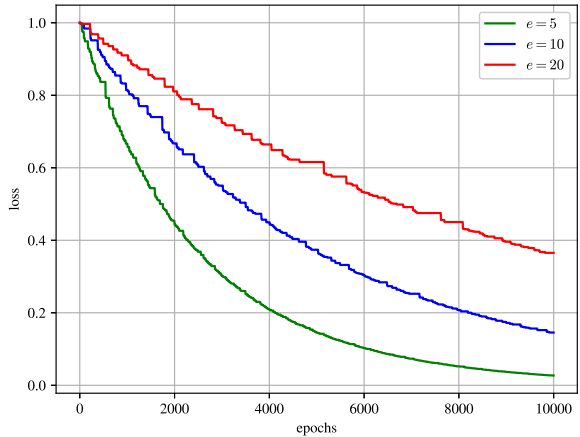}
\label{fig: loss_100}
}\hfill
\subfigure[Staleness for $n=100$.]{
\centering
\includegraphics[scale=0.55]{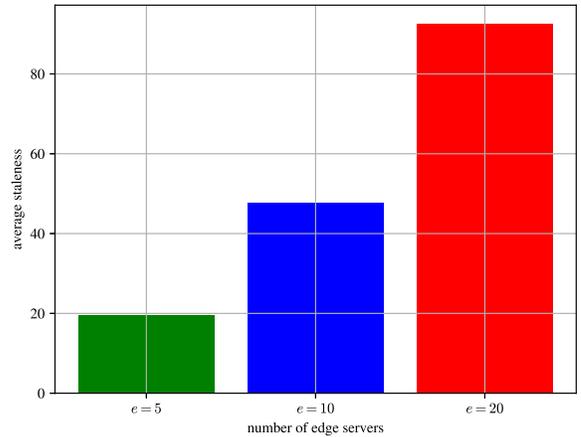}
\label{fig: stale_100}
}
\subfigure[Loss vs. epochs for $n=400$.]{
\centering
\includegraphics[scale=0.55]{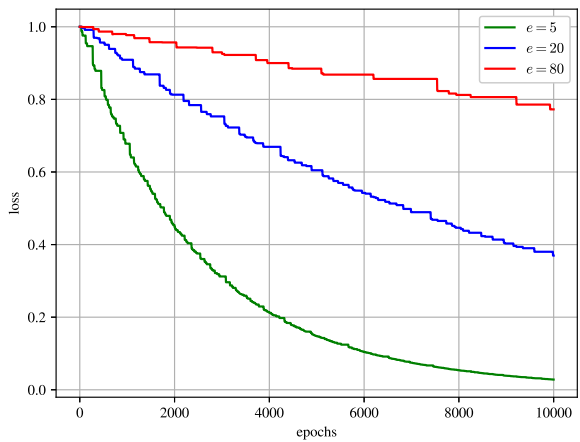}
\label{fig: loss_400}
}\hfill
\subfigure[Staleness for $n=400$.]{
\centering
\includegraphics[scale=0.55]{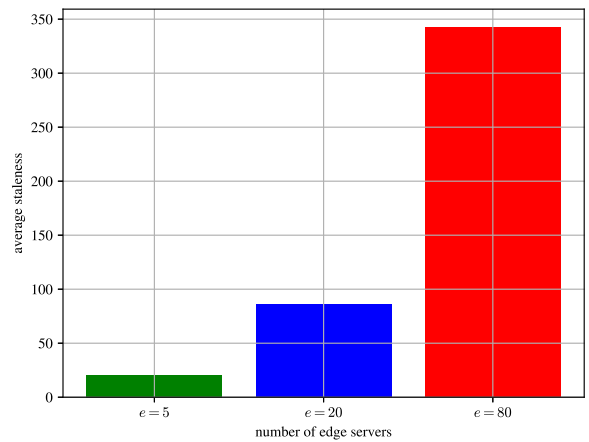}
\label{fig: stale_400}
}
\caption{Loss profiles and staleness comparison for $O(1),O(\sqrt{n})$ and $O(n)$ number of edge servers.}
\label{fig: loss_stale_simulations}
\end{figure*}

\section{Numerical Results}

In this section, we simulate the timely AHFL model for a simple linear regression task. In our experiment, we synthetically generate a dataset, in the same way as \cite{ozfatura2020age, li2018polynomially, buyukates_timely_FL}; divide the dataset into equal parts, give to the clients and perform Algorithm~\ref{algo: AHFL} to numerically calculate the regression loss and the average staleness.

We generate the dataset $D=\{(\boldsymbol{x}_j,y_j)\}$, where $\boldsymbol{x}_j\in\mathbb{R}^{d}$. In our case, $d=100$ and the data points are from the Gaussian mixture distribution $\boldsymbol{x}_j\sim\frac{1}{2}\mathcal{N}\left(\frac{1.5}{d}\boldsymbol{w}^*,\boldsymbol{I}_d\right)+\frac{1}{2}\mathcal{N}\left(-\frac{1.5}{d}\boldsymbol{w}^*,\boldsymbol{I}_d\right)$, where each component of $\boldsymbol{w}^*\in\mathbb{R}^d$ are chosen uniformly from the interval $[0,1]$. The corresponding output is $y_j=\boldsymbol{x}_j^{T}\boldsymbol{w}^*$. Therefore, the overall loss function to minimize is 
\begin{align}\label{loss function formula}
    \mathcal{L}(\boldsymbol{\theta};D)&=\frac{1}{|D|}\sum_{j\in[|D|]}\left(\boldsymbol{x}_j^{T}\boldsymbol{\theta}-y_j\right)^2\\
    &=\frac{1}{|D|}\large||X\boldsymbol{\theta}-\boldsymbol{y}\large||^2_2,
\end{align}
where $X=\left[\boldsymbol{x}_1,\ldots,\boldsymbol{x}_{|D|}\right]^T$ and $\boldsymbol{y}=\left[y_1,\ldots,y_{|D|}\right]^T$ are data and corresponding outputs and $||\cdot||_2$ represents the $l_2$ norm. The loss function achieves its minimum when $\boldsymbol{\theta}=\boldsymbol{w}^*$, since all the error terms in the summation of \eqref{loss function formula} will be $0$. The convergence analysis of \cite{xie2020asynchronous} shows that in order to achieve provable convergence via asynchronous training, this loss function $\mathcal{L}$ must satisfy the criteria of being $L$-smooth and $\mu$-weakly convex. From the formulation in \eqref{loss function formula}, since $\mathcal{L}$ is differentiable, we can write the gradient as
\begin{align}
    \nabla_{\boldsymbol{\theta}}\mathcal{L}(\boldsymbol{\theta};D)=\frac{2}{|D|}X^T(X\boldsymbol{\theta}-\boldsymbol{y}).
\end{align}
Now, for $\boldsymbol{\theta}_1$ and $\boldsymbol{\theta}_2$, we have the following equation
\begin{align}\label{difference_grad}
    \nabla\mathcal{L}(\boldsymbol{\theta}_1;D)-\nabla\mathcal{L}(\boldsymbol{\theta}_2;D)=\frac{2X^TX}{|D|}(\boldsymbol{\theta}_1-\boldsymbol{\theta}_2).
\end{align}
Now, choosing $L=\frac{2}{|D|}||X^TX||_2$, we obtain from \eqref{difference_grad}:
\begin{align}
    ||\nabla\mathcal{L}(\boldsymbol{\theta}_1;D)-\nabla\mathcal{L}(\boldsymbol{\theta}_2;D)||_2=L||\boldsymbol{\theta}_1-\boldsymbol{\theta}_2||_2,
\end{align}
which implies that $\mathcal{L}$ is a $L$-smooth function. Also, since \eqref{loss function formula} is a $l_2$ norm, which is a convex function, it is also $\mu$-weakly convex. Hence, the regression loss function satisfies both the  provable convergence criteria.

In our experiment, $|D|=10,000$. This data is equally distributed among $n$ clients. Therefore, for the $i$th client, $|D_i|=|D|/n$. Each client tries to minimize the modified loss function, defined in \eqref{modified_loss_function}, when they are included in a training cycle. For the simulations, we consider $\rho=0.01$. Also, since the goal of our work is to show the effect of edge servers on convergence, we take both $\alpha$ and $\beta$ as $0.5$, i.e., in each training cycle corresponding to a particular edge server, the edge server waits for half of the clients associated with it to become available with exponential time parameter $\lambda=1$ and half of those clients get to finish their $\tilde{t}$ step gradient descent in time $c=1$ and update the edge server with exponential delay parameter $\tilde{\mu}=1$. In simulations, $\tilde{t}=10$. After aggregating the model at the edge, the cloud server is then updated as \eqref{cloud_update_equation}. We use the following decreasing function
\begin{align}
    \sigma(t-t')=\frac{1}{(t-t')^{0.1}},\quad t>t'
\end{align}
for taking staleness into account. We consider three different orders for the number of edge servers: $e\sim O(1)$, $O(\sqrt{n})$ and $O(n)$. The training process continues for $T=10,000$ epochs, and the obtained results are shown in Fig.~\ref{fig: loss_stale_simulations}. 

First, we consider $n=100$ clients. Each client has $100$ data points drawn from the Gaussian mixture distribution, and they train in the AHFL setting. The number of edge servers is considered as $e=5$, $e=\sqrt{n}=10$ and $e=0.2\times n=20$. From Fig.~\ref{fig: loss_100}, we observe that for $e=5$, the regression loss decreases rapidly, i.e., the client model converges with the fastest rate, whereas the convergence rate slows down as $e$ increases. This is due to the increasing staleness of the system as $e$ increases. For $\alpha$ and $\beta$ as $0.5$, the expected staleness, from \eqref{final staleness}, is $4e-1$, i.e., $19$, $39$ and $79$, respectively. This analytical result is close to the numerically evaluated average staleness, as shown in Fig.~\ref{fig: stale_100}. The effect of this increasing staleness is also observed in Fig.~\ref{fig: loss_100} as the loss profile has more \textit{staircase} decrements due to the individual nodes in the clusters being trained less frequently.

Similar trends are observed for $n=400$. In this case, each client has $25$ data points for training. The number of edge servers is $5$, $20$ and $80$, respectively. Fig.~\ref{fig: loss_400} shows that the $e=5$ is the fastest, as before, and the rate of convergence decreases as $e$ increases. We also observe that for $e=5$ and $20$, the $100$ client setting converges faster than $400$ client setting. This might be due to the fact that with higher number of clients, the data becomes more distributed, which results in slower convergence rate. Finally, the staleness profile, as shown in Fig.~\ref{fig: stale_400}, indicates that with higher number of edge servers, the models become more stale. The average values also match with the analytically calculated mean staleness, which are $19$, $79$ and $319$, respectively.

\section{Conclusion}

We analyzed a timely gossiping scheme in an AHFL model. In this system model, the clients are divided into multiple clusters which are connected to the edge servers which perform local aggregation in a timely manner. In each training cycle, the edge servers send their models to a selected number of clients who become available first. The clients train the model with their locally available data. After completion of the training, the clients send back their trained model to the edge servers. Only a limited number of models, that arrive first, are used for the edge-aggregation. When an edge is updated, it sends its locally aggregated model to a cloud server, which performs asynchronous global aggregation and gives back the new global model to the edge servers with negligible time delay. This model is then used for the next training cycle of the clients. Since each client's trained model is not necessarily aggregated in each training cycle, this model incurs some staleness into the system. With bounded staleness, this timely AHFL training model can converge. We showed that if the number of edge servers is $e=O(1)$, then for a system with dense clients, there is a probabilistic guarantee of convergence of the distributed learning model in a finite time.

\bibliographystyle{unsrt}
\bibliography{reference}

\end{document}